\documentclass[aps,pre,reprint,showpacs,superscriptaddress,groupaddress,showkeys,floatfix]{revtex4-1}
\usepackage{amsmath}
\usepackage{amsfonts}
\usepackage{amssymb}
\usepackage[dvipsnames]{xcolor}
\usepackage{graphicx,color,subeqnarray}
\usepackage{dcolumn}
\usepackage{bm}
\usepackage{srctex}

\begin{document}

\title{Smoluchowski Diffusion Equation for Active Brownian Swimmers}
\author{Francisco J. Sevilla}
\email{fjsevilla@fisica.unam.mx}
\affiliation{Instituto de F\'isica, Universidad Nacional Aut\'onoma de M\'exico,
Apdo.\ Postal 20-364, 01000, M\'exico D.F., Mexico
}

\author{Mario Sandoval}
\email{sem@xanum.uam.mx}
\affiliation{
 Department of Physics, Universidad Autonoma Metropolitana-Iztapalapa, 
 Distrito Federal 09340, Mexico.
}
\date{\today}

\begin{abstract}
We study the free diffusion in two dimensions of active-Brownian swimmers subject to
passive fluctuations on the translational motion and to active fluctuations on the
rotational one. The Smoluchowski equation is derived from a Langevin-like model of
active swimmers, and analytically solved in the long-time regime for arbitrary values
of the P\'eclet number, this allows us to analyze the out-of-equilibrium evolution of
the positions distribution of active particles at all time regimes. Explicit expressions
for the mean-square displacement and for the kurtosis of the probability distribution
function are presented, and the effects of persistence discussed. We show through Brownian
dynamics simulations that our prescription for the mean-square displacement gives the exact
time dependence at all times. The departure of the probability distribution from a Gaussian,
measured by the kurtosis, is also analyzed both analytically and computationally. We find
that for P\'eclet numbers $\lesssim 0.1$, the distance from Gaussian increases as $\sim t^{-2}$
at short times, while it diminishes as $\sim t^{-1}$ in the asymptotic limit.
\end{abstract}


\maketitle

\affiliation{Instituto de F\'isica, Universidad Nacional Aut\'onoma de M\'exico,
Apdo.\ Postal 20-364, 01000, M\'exico D.F., Mexico
}

\affiliation{
 Department of Physics, Universidad Autonoma Metropolitana-Iztapalapa, 
 Distrito Federal 09340, Mexico.
}

\section{Introduction}

The study of self-propelled (active) particles moving at small scales has
received much attention \cite{lovely75, ped, berg93, ishi, how, roman, lp09,
Julicher:2009hw,SevillaPRE2014}. This is so since phenomena in nature like 
motion of plankton, viruses and bacteria, have an important role in many biological
processes as well as in industrial applications, and more generally, because active 
particles are suitable elements of analysis in the context on nonequilibrium 
statistical physics. Moreover, confined matter made by
self-propelled particles has been observed to behave in a very different way
compared to matter conformed by passive ones, mainly due to the out-of-equilibrium
nature of active systems. For example, it has been observed that active particles, 
confined in a box with sharp corners, accumulate more at the corners of the container
rather than spreading uniformly through the box. This effect originates that the 
generated pressure on the walls do not be uniform along the container \cite{Fily}. 
In contrast, matter conformed by passive particles develop a uniform pressure in the
container. It has also been reported that interacting active particles with different
sizes and confined in a box, form clusters leaving empty regions inside the box \cite{agre},
a very different scenario occurs with passive particles since these particles would spread
homogeneously in the container. Thus we notice that the interplay among activity and confinement
may provide novel properties to the latter systems,
hence the necessity of studying them.

From a technological point of view, the study of active particles  is also very relevant.
To mention an example, the bioengineering community is
constructing self-propelled micromachines \cite{abbot,kosa,GaoNano2014} inspired by natural 
swimmers, and with the purpose of making devices able to deliver specialized drugs in precise
region inside our body, or to serve as micromachines able to detect, and
diagnose diseases \cite{chemical2,paxton06,chemical3}. These microrobots,
in the same way as the smallest microorganisms, are subject to
thermal fluctuations or Brownian motion, which is an important issue to take into account,
since thermal fluctuations make the particles to lose their orientation, thus affecting the 
particles net displacement. Most classical literature considering the effect of loss of 
orientation due to Brownian motion and activity on the particles diffusion, has
used a Langevin approach, since it can be considered as a direct way of
finding the particles effective diffusion. Within this approach, isotropic
self-propelled bodies subject to thermal forces \cite{how,hagen, loba} and
anisotropic swimmers \cite{hagen} have been treated in the absence and
presence of external fields \cite{ebbens, van,hagen2,sando2}. A Smoluchowski
approach to study the effective diffusion of active particles in slightly complex environments
has also been undertaken. For example, Pedley \cite{pe0} introduced a continuum model to
calculate the probability density function for a diluted suspension of gyrotactic bacteria. 
Bearon and Pedley \cite{bearon} studied chemotactic bacteria under shear flow and
derived an advection-diffusion equation for cell density. Enculescu and
Stark \cite{encu} studied the sedimentation of active particles due to
gravity and subject to translational and rotational diffusion. Similarly,
Pototsky and Stark \cite{poto} followed a Smoluchowski approach and analyzed
active Brownian particles in two-dimensional traps. Saintillan and Shelley 
\cite{sainti} used a kinetic theory to study pattern formation of
suspensions of self-propelled particles.

Self-propelled particles in nature move in general with
a time-dependent swimming speed, like microorganisms that tend to relax (rest) for
a moment and then to continue swimming \cite{babel}. In this work, we assume that
particles move with a constant swimming speed, which is a simplified model that has
been supported by experimental work \cite{ebbens,cou} and has also been
adopted when studying collective behavior \cite{vic}. Other more complex scenarios
where for example active Brownian particles are subject to external flows, have been
reported and solved following a Langevin approach \cite{sando2}.

In this paper we follow the approach of Smoluchowski and focus on the coarse-grained 
probability density of finding an active Brownian particle---that diffuses translationally
and rotationally in a two-dimensional, unbounded space, and immersed in a steady fluid---at
position $\boldsymbol{x}$ at time $t$ without actually knowing its direction of motion. 
We derive Smoluchowski's equation for such probability density and we solve it analytically. 

We are able to use the Fourier analysis to translate the Fokker-Planck equation 
for active particles, into an infinite set of coupled ordinary differential equations
with a hierarchy similar to the infinite BBGKY (Bogoliubov-Born-Green-Kirkwood-Yvon)
hierarchy that appears in the Boltzmann theory, which in principle can be systematically
solved. We close the hierarchy at the second level and explicitly solve the first hierarchy
equation, which corresponds to the Smoluchowski equation, and provides the probability
density function (p.d.f) we are interested in. We transform back to the coordinate space, 
and we are able to explicitly find the p.d.f. of an active Brownian particle in the short
and long time regimes. The mean square displacement (msd) for a swimmer at short and long times
is also found, and classical results for the enhanced diffusion of an active particle are recovered.
We also find theoretical results for the kurtosis of the swimmer, hence a discussion
concerning the non-Gaussian behavior at intermediate times of an active swimmer
p.d.f. is also offered. Finally and with comparison purposes, Brownian dynamics simulations
were also performed obtaining an excellent agreement among our theoretical and computational
results at short and long times for both, kurtosis and mean-square displacement of the swimmer.

\section{\label{SectModel}The model}

Consider a spherical particle of radius $a$, immersed in a fluid at fixed
temperature $T$, that self-propels in a two-dimensional domain. The particle
is subject to thermal fluctuations (modeled as white noise) in translation
and rotation, that is, $\boldsymbol{\xi }_{T}(t)$ and $\xi _{R}(t)$
respectively, where $\langle \boldsymbol{\xi }_{T}\rangle =\langle \xi
_{R}\rangle =0$, $\langle \xi _{i,T}(t)\xi _{j,T}(s)\rangle =2D_{B}\delta
(t-s)$ and $\langle \xi _{R}(t)\xi _{R}(s)\rangle =2D_{\Omega }\delta (t-s)$%
. Here $D_{B}=k_{B}T/6\pi \eta a$ and $D_{\Omega }$ are respectively, the
translational and rotational diffusivities constants, with $\eta $ the
viscosity of the fluid. 

The particle swimming velocity, $\mathbf{U}_{s}(t)$, is written explicitly
as $U_{s}(t)\hat{\boldsymbol{u}}(t)$, where we denote by $\hat{\boldsymbol{u}%
}(t)=\left[ \cos \varphi (t),\sin \varphi (t)\right] $ ($\varphi (t)$ being
the angle between the direction of motion and the horizontal axis) the
instantaneous unit vector in the direction of swimming, and $U_{s}(t)$ the
instantaneous magnitude of the swimming velocity along $\hat{\boldsymbol{u}}%
(t)$. Each of these quantities may be determined from its own dynamics \cite%
{RomanczukPRL2011}, and Langevin equations for each may be written. Here we
consider the case of a faster dynamics for the swimming speed such that $%
U_{s}(t)=U_{0}=$const. In this way, the dynamics of this active Brownian
particle, subject to passive-translational and active-rotational noise, is
determined by its position $\boldsymbol{x}(t)$ and its direction of motion $%
\hat{\boldsymbol{u}}(t)$, computed from $\varphi(t)$, that obey the following 
Langevin equations 
\begin{subequations}
\label{modelo}
\begin{align}
\frac{d{}}{dt}\boldsymbol{x}(t)& =U_{0}\,\hat{\boldsymbol{u}}(t)+\boldsymbol{%
\xi}_{T}(t),  \label{L1} \\
\frac{d}{dt}\varphi (t)& =\xi _{R}(t).  \label{L2}
\end{align}
\end{subequations}
In this way, the temporal evolution of the particle's position [Eq.\eqref{L1}] 
is thus determined by two independent stochastic effects, one that corresponds to
translational fluctuations due to the environmental noise, the other, to the swimmer's
velocity whose orientation is subject to active fluctuations [Eq. \eqref{L2}].
These same equations have been considered in Ref.\cite{ZhengPRE2013}
to model the motion of spherical Platinum-silica Janus particles in a
solution of water and H$_{2}$O$_{2}$. 

From now on, we use $D_{\Omega}^{-1}$ and $U_{0}D_{\Omega}^{-1}$
as time and length scales respectively, such that $\widetilde{D}_{B}\equiv D_{B}
D_{\Omega}/U_{0}^{2}$ is the only free, dimensionless parameter of our analysis
which coincides with the inverse of the so called P\'eclet number, which in this case 
corresponds to the ratio of the advective transport coefficient, $U_{0}^{2}/D_{\Omega},$
to the translational diffusion transport coefficient $D_{B}$. For the Janus particles
studied by Pallaci \emph{et al.} in Ref. \cite{PalacciPRL2010}, we have that 
$D_{\Omega}^{-1}\approx0.9$ s$^{-1}$, $D_{B}\approx0.34$ $\mu$m$^{2}$/s and swimming 
speeds between 0.3 and 3.3 $\mu$m/s. These values give $\widetilde{D}_{B}$
between 0.035 and 4.2. Numerical results, on the other hand, are obtained by integrating
Eqs. \eqref{modelo} using the Euler-Crome explicit scheme with a time-step of 0.005.
\begin{figure}
 \includegraphics[width=\columnwidth]{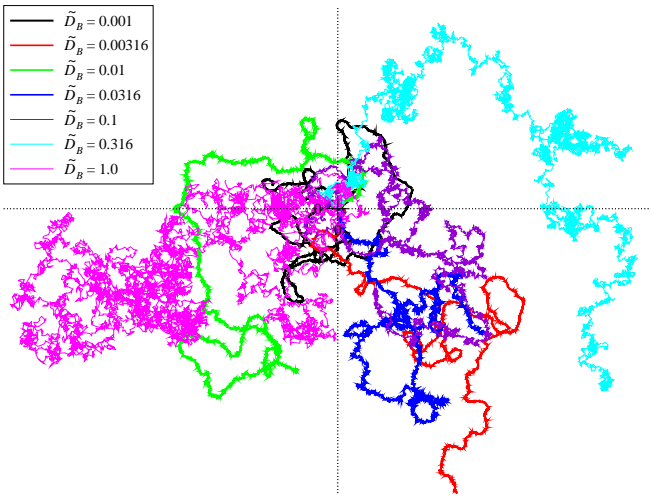}
 \caption{(Color online) Single particle trajectories for different values of 
 $\widetilde{D}_{B}=0.001,\, 0.00316,\, 0.01,\, 0.0316,\, 0.1,\, 0.316,\, 1.0$.
 Data is generated by solving Eq.~\eqref{modelo} during $10^{4}$ time steps, 
 the initial position is chosen at the origin while the initial orientation 
 is drawn from a uniform distribution in $[0,2\pi)$.}
 \label{Trajectories}
\end{figure}
In Fig. \ref{Trajectories} seven single-particle trajectories are
presented to show the effect of varying $\widetilde{D}_{B}.$ The persistence effects
are conspicuous for small values of $\widetilde{D}_{B}$ (thick curves). 

An equation for the one particle probability
density $P({\boldsymbol{x}},\varphi ,t)\equiv \langle \delta ({\boldsymbol{x}%
}-{\boldsymbol{x}}(t))\delta (\varphi -\varphi (t))\rangle$ can be derived by
differentiating this $P({\boldsymbol{x}},\varphi ,t)$ with respect to time,
namely
\begin{multline}
\frac{\partial }{\partial t}P({\boldsymbol{x}},\varphi ,t)+U_{0}\hat{%
\boldsymbol{u}}\cdot \nabla P({\boldsymbol{x}},\varphi ,t)= \\
-\frac{\partial }{\partial \varphi }\langle \xi_{R} (t)\delta ({\boldsymbol{x}}-{%
\boldsymbol{x}}(t))\delta (\varphi -\varphi (t))\rangle\\
-\nabla\cdot\langle \boldsymbol{\xi}_{T}(t)\delta ({\boldsymbol{x}}-{%
\boldsymbol{x}}(t))\delta (\varphi -\varphi (t))\rangle,
\end{multline}%
where $\langle \cdot \rangle $ denotes the average over translational and
rotational noise realizations and $\nabla =(\partial /\partial x,\partial /\partial y).$ 
By using Novikov's theorem, we obtain the
following Fokker-Planck equation for an active Brownian particle 
\begin{multline}
\frac{\partial }{\partial t}P({\boldsymbol{x}},\varphi ,t)+U_{0}\,\hat{%
\boldsymbol{u}}\cdot \nabla P({\boldsymbol{x}},\varphi ,t) = \\ D_{B}\nabla
^{2}P({\boldsymbol{x}},\varphi ,t) 
+D_{\Omega }\frac{\partial ^{2}}{\partial \varphi ^{2}}P({\boldsymbol{x}}%
,\varphi ,t). \label{smo}
\end{multline}%
We now start to analytically solve Eq.~(\ref{smo}). To do so, we apply the
Fourier transform to Eq. (\ref{smo}) and we obtain
\begin{multline}
\frac{\partial }{\partial t}\widetilde{P}({\boldsymbol{k}},\varphi
,t)+iU_{0}\,\hat{\boldsymbol{u}}\cdot {\boldsymbol{k}}\widetilde{P}({%
\boldsymbol{k}},\varphi ,t) =\\-D_{B}{\boldsymbol{k}}^{2}\widetilde{P}({%
\boldsymbol{k}},\varphi ,t)  
+D_{\Omega }\frac{\partial ^{2}}{\partial \varphi ^{2}}\widetilde{P}({%
\boldsymbol{k}},\varphi ,t),\label{trans}
\end{multline}
where
\begin{equation}
\widetilde{P}({\boldsymbol{k}},\varphi ,t)=(2\pi )^{-1}\int d^{2}\boldsymbol{%
x}\,e^{i\boldsymbol{k}\cdot \boldsymbol{x}}\,P(\boldsymbol{x},\varphi ,t),
\label{fourier}
\end{equation}
and ${\boldsymbol{k}}=(k_{x},k_{y})$ denotes the Fourier wave-vector.
For $U_{0}=0$, Eq. \eqref{trans} has as solution the set of eigenfunctions 
$\{e^{-(D_{B}k^{2}+D_{\Omega}n^{2})t}e^{in\varphi}\}$ with $n$ an integer.
Thus we expand $\widetilde{P}(\boldsymbol{k},\varphi ,t)$ on this set,
expressly
\begin{equation}
\widetilde{P}({\boldsymbol{k}},\varphi ,t)=\frac{1}{2\pi}\sum\limits_{n=-\infty}
^{\infty }\widetilde{p}_{n}({\boldsymbol{k}}%
,t)e^{-(D_{B}k^{2}+D_{\Omega}n^{2})t}e^{in\varphi},  \label{s1}
\end{equation}
with $k=\vert\boldsymbol{k}\vert$. This expansion
allows to separate the effects of translational diffusion contained in the prefactor
$e^{-D_{B}k^{2}t}$ from the effects of rotational diffusion due to active fluctuations.
 
The coefficients of the expansion are given by
\begin{equation}
\widetilde{p}_{n}({\boldsymbol{k}},t)=e^{(D_{B}k^{2}+D_{\Omega}n^{2})t}\int_{0}^{2\pi}d\varphi
\, \widetilde{P}({\boldsymbol{k}},\varphi ,t)e^{-in\varphi }.   \label{inv}
\end{equation}
After substituting Eq.~\eqref{s1} in Eq.~\eqref{trans} and using th orthogonality of the
eigenfunctions we get the following set of coupled ordinary differential
equations for the $n$-th coefficient of the expansion $\widetilde{p}_{n}({%
\boldsymbol{k}},t),$ namely%
\begin{multline}\label{rec}
\frac{d}{dt}\widetilde{p}_{n} =-\frac{U_{0}}{2}ik\, e^{-D_{\Omega }t}\left[
e^{2nD_{\Omega }t}e^{-i\theta}\,\widetilde{p}_{n-1}+\right. 
\\
\left. e^{-2nD_{\Omega }t}e^{i\theta}\,\widetilde{p}%
_{n+1}\right] . 
\end{multline}
Note that we have introduced the quantities $k_{x}\pm ik_{y}=ke^{\pm i\theta }$. 

Our interest lies on $P_{0}(\boldsymbol{x},t)=\int_{0}^{2\pi}d\varphi\, P(\boldsymbol{x}
,\varphi,t)$, which is given by the inverse Fourier transform of $\widetilde{P}_{0}
(\boldsymbol{k},t)=e^{-D_{B}k^{2}t}\widetilde{p}_{0}
(\boldsymbol{k},t)$ as can be checked from Eq.~\eqref{inv}.
Let us find $\widetilde{p}_{0}$ by solving Eq.~(\ref{rec}) when the initial condition $\widetilde{P}%
_{n}(\boldsymbol{k},0)=\widetilde{p}_{n}(\boldsymbol{k},0)=\delta
_{n,0}/2\pi$ is considered. This initial condition corresponds to
the case when the particle departs from the origin in a random direction
of motion drawn from a uniform distribution in $[0,2\pi)$ and is equivalent to the initial condition $P({\boldsymbol{x}%
},\varphi ,0)=\delta ^{(2)}({\boldsymbol{x}})/2\pi $, where $\delta _{n,m}$
and $\delta ^{(2)}({\boldsymbol{x}})$ denote, respectively, the Kronecker
delta and the delta function.

We first deal with the solution of Eq.~(\ref{rec}) for long times. To do so,
we consider only the first three Fourier coefficients $\widetilde{p}_{n},$ i.e., those 
with $n=0,\,\pm 1$, later on it will be shown that the other coefficients do not
contribute in this regime. From Eq. \eqref{rec} the three first Fourier coefficients
satisfy
\begin{subequations}
\begin{align}
\frac{d}{dt}\widetilde{p}_{0}=-\frac{U_{0}}{2}ik\,e^{-D_{\Omega }t}\left[
e^{-i\theta }\widetilde{p}_{-1}+e^{i\theta }\widetilde{p}_{1}\right] ,
\label{rec1}\\
{\frac{d}{dt}\widetilde{p}_{\pm 1}}=-\frac{U_{0}}{2}ik\,{\left[ e^{D_{\Omega
}t}e^{\mp i\theta }\widetilde{p}_{0}+e^{-3D_{\Omega }t}e^{\pm i\theta }%
\widetilde{p}_{\pm 2}\right] ,}  \label{recu2}
\end{align}
\end{subequations}
which after some algebra are combined into a single equation for $\widetilde{p}_{0}$,
that is
\begin{multline}\label{lead0}
\frac{d^{2}}{dt^{2}}\widetilde{p}_{0}+D_{\Omega }\frac{d}{dt}\widetilde{p}%
_{0}=-\frac{U_{0}^{2}}{2}k^{2}\widetilde{p}_{0}\\
-\frac{U_{0}^{2}}{4}k^{2}e^{-4D_{\Omega}t}\left(e^{2i\theta}\widetilde{p}_{2}
+e^{-2i\theta}\widetilde{p}_{-2}\right).  
\end{multline}
At this stage we argue that in the long time regime we may neglect the second term
in the rhs of Eq.~(\ref{lead0}). This approximation leads to the telegrapher's
equation 
\begin{equation}
\frac{d^{2}}{dt^{2}}\widetilde{p}_{0}+D_{\Omega }\frac{d}{dt}\widetilde{p}%
_{0}=-\frac{U_{0}^{2}}{2}k^{2}\widetilde{p}_{0},  \label{lead}
\end{equation}%
which is rotationally symmetric in the $\boldsymbol{k}$-space and therefore,
giving rise to rotationally symmetric solutions in spatial coordinates if
initial conditions with the same symmetry are chosen. One may check that 
Eq.~\eqref{lead} has the solution 
\begin{equation}
\widetilde{p}_{0}(\boldsymbol{k},t)=\widetilde{p}_{0}(\boldsymbol{k}%
,0)\,e^{-D_{\Omega }t/2}\left[ \frac{D_{\Omega }}{2\omega _{k}}\sin \omega
_{k}t+\cos \omega _{k}t\right] ,  \label{lead2}
\end{equation}%
with $\omega _{k}^{2}\equiv U_{0}^{2}k^{2}/2-D_{\Omega }^{2}/4$.
Therefore, by using this result one can explicitly write
that $\widetilde{P}_{0}(\boldsymbol{k},t)=e^{-D_{B}{k}^{2}t}\widetilde{p}%
_{0}(\boldsymbol{k},t),$ with $e^{-D_{B}{k}^{2}t}$ being the translational
diffusion propagator. The solution in spatial coordinates, $P_{0}(\boldsymbol{x},t)$,
is obtained after taking the inverse Fourier transform of $\widetilde{P}_{0}(\boldsymbol{k},t)$,
which is given by the convolution of the translational
propagator with the probability distribution that retain the effects of active diffusion,
i.e.,
\begin{equation}
P_{0}(\boldsymbol{x},t)=\int d^{2}\boldsymbol{x}^{\prime }G\left(\boldsymbol{x}-\boldsymbol{x}
^{\prime},t\right)
p_{0}(\boldsymbol{x}^{\prime },t),  \label{prob1}
\end{equation}%
where $p_{0}(\boldsymbol{x},t)$ is the inverse Fourier transform of $%
\widetilde{p}_{0}(\boldsymbol{k},t)$ and 
\begin{equation}
G\left(\boldsymbol{x},t\right)=\frac{e^{-\boldsymbol{x}^{2}/4D_{B}t}}{4\pi D_{B}t},
\end{equation}
is the Gaussian propagator of translational diffusion in two dimensions.

In the asymptotic limit ($t\rightarrow \infty $), at which the coherent
wave-like behavior related to the second order time derivative in Eq.~\eqref{lead} can be neglected
(mainly due to the random dispersion of the particles direction of motion), $p_{0}(\boldsymbol{x},t)$ 
tends to a Gaussian distribution with diffusion constant $U_{0}^{2}/2D_{\Omega}$, namely
\begin{equation}
p_{0}(\boldsymbol{x},t) \xrightarrow[t\rightarrow\infty]{}\frac{e^{-\boldsymbol{x}^{2}/4\left(
U_{0}^{2}/2D_{\Omega }\right) t}}{4\pi \left( U_{0}^{2}/2D_{\Omega }\right) t%
}.  \label{prob2}
\end{equation}
Substitution of this last expression into Eq.~(\ref{prob1}) and performing the
integral, one gets in the long-time regime 
\begin{equation}
P_{0}(\boldsymbol{x},t)=\frac{e^{-\boldsymbol{x}^{2}/4\left(
D_{B}+U_{0}^{2}/2D_{\Omega }\right) t}}{4\pi \left(
D_{B}+U_{0}^{2}/2D_{\Omega }\right) t},  \label{proba}
\end{equation}
from which the classical effective diffusion constant $D=D_{B}+U_{0}^{2}/2D_{\Omega }$
is deduced \cite{PalacciPRL2010}.

In the short time regime $D_{\Omega}t\ll1$, we approximate $\widetilde{p}_{0}(\boldsymbol{k},t)
\approx\widetilde{p}_{0}(\boldsymbol{k},0)=(2\pi)^{-1}$ and the p.d.f. in spatial coordinates
corresponds to the Gaussian
\begin{equation}\label{GaussianShort}
P_{0}(\boldsymbol{x},t)\approx\frac{1}{2\pi}\frac{e^{-\boldsymbol{x}^{2}/4D_{B}t}}{4\pi D_{B}t}.
\end{equation}
Though it is a difficult task to obtain from Eq.~\eqref{prob1} the explicit dependence on $\boldsymbol{x}$ 
and $t$ of $P_{0}$, a formula in terms of a series expansion of the operator $\nabla^{2}$ applied
to $G(\boldsymbol{x},t)$ can be derived, to say
\begin{multline}\label{ProbExact}
P_{0}(\boldsymbol{x},t)=\frac{1}{2\pi}e^{-D_{\Omega}t/2}\sum_{s=0}^{\infty}\frac{1}{(2s)!}\left(1+\frac{D_{\Omega}t}{2}\frac{1}{2s+1}\right)\times\\
\left(\frac{D_{\Omega}t}{2}\right)^{2s}\left(1+\frac{2U_{0}^{2}}{D_{\Omega}^{2}}\nabla^{2}\right)^{s}\, G(\boldsymbol{x},t),
\end{multline}
Note that the last formula can be rewritten in terms of $\partial_{t}$ instead of $\nabla^{2}$ 
by using that $\nabla^{2}G=D_{B}^{-1}\partial_{t}G.$ 

\begin{figure}
\includegraphics[width=\columnwidth]{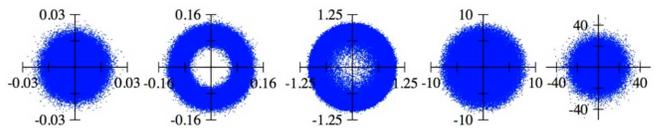}
\caption{(Color online) Snapshots of the positions of $10^{5}$ particles taken at different times, 
namely from left to right, $D_{\Omega}t=0.01,\, 0.1,\, 1.0,\, 10,$ and 100, for an inverse P\'eclet number
$\widetilde{D}_{B}=0.001$. Notice the formation of a rotationally symmetric ring-like structure (second 
panel from left) that develops from a Gaussian distribution (far left panel) due to the effects of persistence
in the motion of the particles. The structure fades out as time
passes (third and fourth panels from left) to reach a Gaussian distribution at long times (far right panel)
.}
\label{fig_distributions}
\end{figure}

In figure \ref{fig_distributions} snapshots of the positions of $10^{5}$ particles taken 
at different times from numerical simulations, and for the inverse P\'eclet number 
$\widetilde{D}_{B}=0.001$, are shown. The first panel from left ($D_{\Omega}t=0.01$),
corresponds to a p.d.f. close to the Gaussian given by Eq.~\eqref{GaussianShort}. Note
the formation of a rotationally symmetric ring-like structure at the intermediate time
$D_{\Omega}t=0.1$ (second panel from left), mainly due to the persistence effects on the particles'
motion. The structure fades out with time (third and fourth panels from left) and starts 
turning into the Gaussian distribution [Eq.~\eqref{proba}], as time passes (far right panel
for $D_{\Omega}t=100$). We must mention that in spite of the good agreement in the short and
asymptotic limits given by \eqref{ProbExact}, this solution does not provide an adequate
description of the particles distribution in the intermediate time regime for all values
of $\widetilde{D}_{B}$, particularly for the small ones, for which the effects of persistence
are conspicuously prominent. The reason of this failure lies, not in the general solution \eqref{prob1}
but in the approximated equation \eqref{lead}, which corresponds to the two dimensional
telegrapher's. The solution to this equation [\eqref{lead2}] can not be interpreted as a p.d.f. 
since becomes negative at short times \cite{PorraPRE1997}. This undesired feature, results from the wake effect,
characteristic of the solution of the two dimensional wave equation \cite{SevillaPRE2014}. Notwithstanding this and as
is shown later on, the mean-square displacement computed from our approximation coincides with the
exact one computed from numerical simulations, at all time regimes.

\section{Mean-Square Displacement}
What are the effects of self-propulsion on the diffusive behavior of the system is a question
that has been addressed in several experimental and theoretical studies, and the msd, being a
measure of the covered space as a function of time by the random particle, has been of physical
relevance in both contexts. Indeed, the msd is obtained in many experimental situations that
consider active particles \cite{SelmecziBiophysJ2005,hagen,ZhengPRE2013}, hence a way of validating
existing theoretical approaches. On the other hand, it is known that the diffusion coefficient of
active particles depends on the particle density, due mainly to excluded volume effects of the particles.
Here we assume an enough diluted system to neglect those effects.  

In what follows, an exact analytical expression for the msd will be obtained. From Eq. (\ref{lead})
one can show that in the coordinate space $P_{0}(\boldsymbol{x},t)$ satisfies
\begin{multline}
\frac{\partial ^{2}}{\partial t^{2}}P_{0}+D_{\Omega }\frac{\partial }{%
\partial t}P_{0}=\left(\frac{U_{0}^{2}}{2}+D_{B}D_{\Omega}\right)\nabla
^{2}P_{0} \\
+D_{B}\left(2\frac{\partial}{\partial t}-D_{B}\nabla^{2}\right)%
\nabla^{2}P_{0}.  \label{smop0}
\end{multline}
If last equation is multiplied by $\boldsymbol{x}^{2}$ an integrated
over the whole space, we obtain for the msd the equation
\begin{equation}
\frac{d^{2}}{dt^{2}}\langle \boldsymbol{x}^{2}(t)\rangle +D_{\Omega }\frac{d%
}{dt}\langle \boldsymbol{x}^{2}(t)\rangle =2U_{0}^{2}+4D_{B}D_{\Omega },
\label{ode}
\end{equation}%
whose solution can be easily found, namely 
\begin{multline}  \label{MSD_analytical}
\langle \boldsymbol{x}^{2}(t)\rangle =4\frac{D_{B}}{D_{\Omega }}\left(
1-e^{-D_{\Omega} t}\right)\\
+\frac{2\left(2D_{B}D_{\Omega}+U_{0}^{2}\right)}{D_{\Omega}^{2}}\left[D_{\Omega}t-\left( 1-e^{-D_{\Omega} t}\right) \right],  
\end{multline}%
where we have used that $\left( d/dt\right) \left. \langle \boldsymbol{x}%
^{2}(t)\rangle \right\vert _{t=0}=4D_{B}$ in this case. 

The linear dependence on time of the msd, expected in the Gaussian regime,
is checked straightforwardly from Eq.~\eqref{MSD_analytical}. In the long-time
regime we get
\begin{equation}
\langle \boldsymbol{x}^{2}(t)\rangle \xrightarrow[D_{\Omega}t\rightarrow\infty]{} 4\left( D_{B}+\frac{U_{0}^{2}}{%
2D_{\Omega }}\right) t,  \label{meanfree}
\end{equation}%
which is a classical result indicating that the effective diffusion of a
self-propelled particle is enhanced by its activity \cite{berg93}. In the
opposite limit.
\begin{equation}
\left\langle \boldsymbol{x}^{2}(t)\right\rangle\xrightarrow[D_{\Omega}t\rightarrow0]{}4D_{B}t,  \label{mso}
\end{equation}%
which once again the latter equation shows, the linear behavior with time of the msd.
\begin{figure}
\includegraphics[width=\columnwidth]{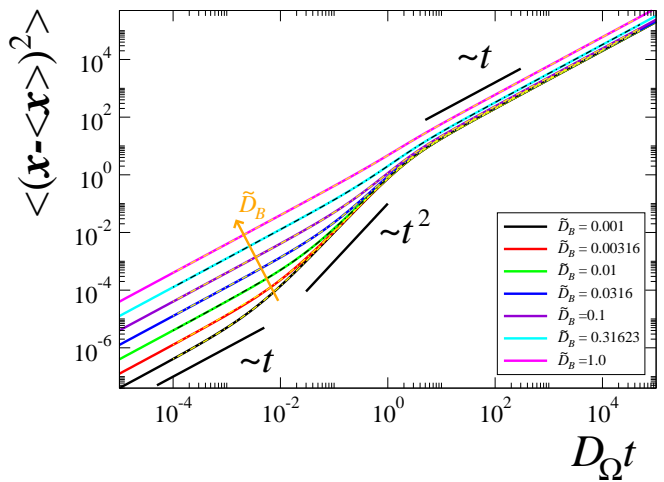}
\caption{(Color online) Mean-square displacement in units of $
U_{0}^{2}/D_{\Omega}^{2}$ as function of the dimensionless time $D_{\Omega}t$	
for different values of the inverse P\'eclet number $\widetilde{D}_{B}$, 
namely 0.001, 0.0031, 0.01, 0.0316, 0.1, 0.3162, 1.0.
Solid lines correspond to the analytical expression given by Eq.~\eqref{MSD_analytical}, while the thin, 
dash-dotted lines correspond to the results obtained from numerical simulations (see text).
}
\label{Fig_msd}
\end{figure}
\subsection{Comparison of the mean-square displacement}

In this subsection we compare the results obtained for the msd at short and
long times using Eq.~(\ref{smop0}) with results obtained following our new
Fourier approach.

For long times and with the explicit expression for $P_{0}(\boldsymbol{x},t)$ [Eq. (\ref{proba})], we use
the definition of the mean-square displacement, $\langle \boldsymbol{x}%
^{2}(t)\rangle =\int d^{2}\boldsymbol{x}\,\boldsymbol{x}^{2}P_{0}(%
\boldsymbol{x},t)$ to easily recover Eq.~(\ref{meanfree}).
For short times, we expand $\widetilde{P}_{0}(\boldsymbol{k},t)$ in powers of $k$, and we keep only the quadratic
terms in $k$ for the reason that will be clear immediately afterwards, we have
\begin{equation}
\widetilde{P}_{0}(\boldsymbol{k},t)=1+\left( \frac{D_{\Omega }}{2}%
-D_{B}k^{2}\right) t-\left( \frac{D_{B}D_{\Omega }k^{2}}{2}+\frac{\omega
_{k}^{2}}{2}\right) t^{2}.  \label{short}
\end{equation}%
We now use that 
\begin{equation}
\left\langle \boldsymbol{x}^{2}(t)\right\rangle =-\nabla _{{\boldsymbol{k}}}^{2}\left. \widetilde{P}_{0}(%
\boldsymbol{k},t)\right\vert _{{\boldsymbol{k}}=0}=-\frac{1}{k}\frac{\partial }{\partial k}k%
\frac{\partial }{\partial k}\left. \widetilde{P}_{0}(\boldsymbol{k}%
,t)\right\vert _{k=0}, \label{rule}
\end{equation}%
hence, at short times, the msd of the active particle is finally given by
\begin{equation}
\left\langle \boldsymbol{x}^{2}(t)\right\rangle\approx 4D_{B}t+\left(2D_{B}D_{\Omega
}+U_{0}^{2}\right)t^{2},  \label{ms}
\end{equation}%
which shows linear and quadratic behavior of the msd. In the limit $%
t\rightarrow 0$, the above equation reduces to Eq.~(\ref{mso}). Thus we have arrived to the same results by two distinct
methods, namely, one method that obtained the explicit expression for the
probability density and used it, and the second method that extracted
information from Eq.~(\ref{smop0}) without solving it.

In order to validate our theoretical findings, we have also
performed Brownian dynamics simulations. Fig.~\ref{Fig_msd} shows a comparison among our theoretical
prediction for any time given by Eq.~(\ref{MSD_analytical}) (solid lines),
and our Brownian dynamics simulations (dashed lines) implemented to solve
Eqs.~\eqref{modelo} and averaging over 10$^{5}$ trajectories. Each
plotted line corresponds to different values of $\widetilde{D}_{B}$.
We observe and excellent agreement among theory
and Brownian simulations. Additionally, Fig.~\ref{Fig_msd} indicates that a
transition of the msd from a linear to a quadratic behavior is not always
the case. For large values of $\widetilde{D}_{B}$ the linear
behavior over time dominates, whereas for $\widetilde{D}_{B}$
smaller than $0.3$, a transition starts to occur. The absence of a
transition from a linear to a quadratic behavior has also been reported by
ten Hagen \textit{et al.} \cite{hagen}. They suggest that at short times,
the msd of a self-propelled particle is affected
by the initial orientation and the magnitude of its force propulsion that make that a transition  occur or not. 

\section{Kurtosis}

Once we have obtained approximately $P_{0}(\boldsymbol{x},t)$ from Eq.
(\ref{smo}), we wish to characterize its departure from a Gaussian p.d.f. as a function
of time. This quantity has been measured experimentally is systems of Janus particles
To this purpose we calculate its kurtosis $\kappa$, given explicitly by \cite{Mardia74p115}
\begin{equation}
\kappa =\left\langle \left[ (\boldsymbol{x}-\langle \boldsymbol{x}\rangle
)^{T}\Sigma ^{-1}(\boldsymbol{x}-\langle \boldsymbol{x}\rangle )\right]
^{2}\right\rangle ,  \label{kur}
\end{equation}%
where $\boldsymbol{x}^{T}$ denotes the transpose of the vector $\boldsymbol{x%
}$ and $\Sigma $ is the $2\times2$ matrix defined by the average of the dyadic product $%
(\boldsymbol{x}-\langle \boldsymbol{x}\rangle )^{T}\cdot (\boldsymbol{x}%
 -\langle \boldsymbol{x}\rangle ).$ In addition, it can be shown that for circularly symmetric
distributions, as the ones considered in the present study, Eq. (\ref{kur})
reduces to%
\begin{equation}
\kappa =4\frac{\langle \boldsymbol{x}^{4}(t)\rangle _{r}}{\langle 
\boldsymbol{x}^{2}(t)\rangle _{r}^{2}},  \label{ksim}
\end{equation}
here $\langle \boldsymbol{\cdot}\rangle _{r}$ stands for the radial
average over the radial probability density distribution, \ namely $\langle 
\boldsymbol{\cdot }\rangle _{rad}=\int_{0}^{\infty }dr\, rP(r)(
\cdot).$ The analytical results obtained for the kurtosis will also be compared with
those obtained from numerical simulations.

\begin{figure}
\includegraphics[width=\columnwidth]{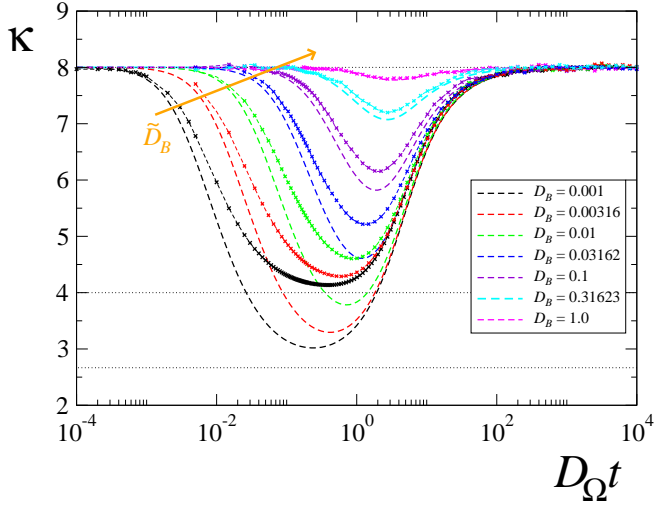}
\caption{(Color online) Kurtosis of the particles distribution as function
of the dimensionless time $D_{\Omega}t$ for different values of $\widetilde{D}_{B}$,
namely 0.001, 0.0031, 0.01,
0.0316, 0.1, 0.3162, 1.0. Dashed lines correspond to the exact analytical
expression given by the quotient of Eq.~\eqref{4thMoment} and 
Eq.~\eqref{MSD_analytical}, while the symbols mark the values of the
kurtosis obtained from numerical simulations. Thin-dotted lines mark the
values 8, 4, 8/3, that correspond respectively in two dimensions to: a
Gaussian distribution, a ring-like distribution and wave-like distribution 
\protect\cite{SevillaPRE2014}.}
\label{Figkur}
\end{figure}

As we can see, Eq.~(\ref{ksim}) requires the calculation of the fourth moment.
After certain algebraic steps  (see appendix \ref{appendixB}), we explicitly
find that the fourth moment is given by
\begin{widetext}
\begin{multline}
\langle \boldsymbol{x}^{4}(t)\rangle _{r}=2^{5}\frac{U_{0}^{4}}{D_{\Omega}^{4}}e^{-D_{\Omega}t/2}\left\{\left[-3\frac{D_{\Omega}t}{2}+\left(\frac{D_{\Omega}t}{2}\right)^{2}\left(1+4\widetilde{D}_{B}(1+\widetilde{D}_{B})\right) \right] \cosh\left(\frac{D_{\Omega}t}{2}\right)+\right. \\
\left.\left[3-\frac{D_{\Omega}t}{2}(1+4\widetilde{D}_{B})+\left(\frac{D_{\Omega}t}{2}\right)^{2} (1+4\widetilde{D}_{B}(1+\widetilde{D}_{B}))\right] \sinh\left(\frac{D_{\Omega}t}{2}\right)\right\}. 
\label{4thMoment}
\end{multline}
\end{widetext}
The latter analytical expression is used in Fig.~\ref{Figkur} to plot the kurtosis,
to be precise, the ratio of Eq.~\eqref{4thMoment} to the square of Eq.~\eqref{MSD_analytical}.
In this figure, kurtosis is plotted as a function of the dimensionless time $D_{\Omega}t$
for different values of $\widetilde{D}_{B}$ (dashed lines), while the corresponding exact
results from Brownian dynamics simulations are shown in symbols. Note that for these last 
ones, $4\le\kappa\le8$.

We observe that the kurtosis shows a clear non-monotonic behavior for $\widetilde{D}_{B}\lesssim1$,
namely, it starts to diminish from $\kappa=8$ as time passes, until it reaches a minimum at a time around
$D_{\Omega}^{-1}$ that depends on the inverse of the P\'eclet number $\widetilde{D}_{B}$. This behavior
has been observed in experiments with Janus particles in three dimensions \cite{ZhengPRE2013} and
theoretically predicted in one quasi-one-dimensional channels \cite{hage2}.
In the limit of vanishing translational diffusion, the kurtosis is a monotonic increasing function of 
time, with $\kappa=4$ as its minimum value \cite{SevillaPRE2014}.
At later times the kurtosis starts to increase reaching the value 8 in the asymptotic limit. This
behavior becomes less and less evident as the translational diffusion coefficient surpasses the
effective diffusion coefficient that originates in the rotational diffusion, $U_{0}^{2}/D_{\Omega}.$  
Due to this effect, we note a better agreement between our analytical approximation and our simulation
results, as $\widetilde{D}_{B}$ is increased. On the other hand, the persistence effects induced by
orientational correlations are revealed in the diminishing regime of the kurtosis, as is shown in Fig.
\ref{Figkur} for the cases in which $\widetilde{D}_{B}\lesssim1$ (see Table \ref{table} for the values
of $\kappa$ corresponding to the snapshots shown in Fig. \ref{fig_distributions}). In this regime the orientational
correlations surpass the effects of translational noise hence allowing the particles to move persistently
outwardly from the origin and giving rise to circularly-symmetric, ring-like distributions as the one shown
in Fig.~\ref{fig_distributions} (second from left) at $D_{\Omega}t=0.1$ for $\widetilde{D}_{B}=0.001$.
Afterwards, this ring-like distributions turns into a Gaussian distribution as the orientational correlations
die out.  
Notice that the discrepancy between the results given by our analytical approximation and the exact 
results lies on the wave-like effects given by the second-order-time-derivative term of Eq.~\eqref{lead}
which, as shown in \cite{SevillaPRE2014}, gives $8/3\simeq2.6667$ as the value for the kurtosis in the 
vanishing translational diffusion. For $\widetilde{D}_{B}=0.001$, the minimum of the exact kurtosis
is about $4.1352$ at $t\approx0.39D_{\Omega}^{-1}$  while from our analytical approximate
$\kappa\approx3.0184$ at $t\approx0.239D_{\Omega}^{-1}$.

In the next subsections we find asymptotic expression for the kurtosis for both short and long times that
confirm the Gaussianity of the distribution function for these time regimes.
\begin{table}
\caption{\label{table} Kurtosis values at times $D_{\Omega}t$ that are close to those values chosen in Fig. 
\ref{fig_distributions} ($\widetilde{D}_{B}=0.001$).}
\begin{ruledtabular}
\begin{tabular}{lddddd}
$D_{\Omega}t$ & 0.01 & 0.1 & 1.01 & 10.245 & 100.08\\
$\kappa$ & 5.9598 & 4.3111 & 4.2729 & 6.7621 & 7.8713 \\
\end{tabular}
\end{ruledtabular}
\end{table}

\subsection{Kurtosis at short times}
In ordet to find $\kappa $ for $t\rightarrow 0,$ we expand $\widetilde{P}_{0}(\boldsymbol{k},t)$
in a power series in $k$, use the fact that $\omega _{k}^{2}\equiv
U_{0}^{2}k^{2}/2-D_{\Omega }^{2}/4$, and keep terms only of order $k^{4}$ to
get%
\begin{eqnarray}
\widetilde{P}_{0}(\boldsymbol{k},t) &\approx &k^{4}\left[ \frac{D_{\Omega }}{%
2}\frac{c^{4}t^{5}}{5!}+\frac{c^{4}t^{4}}{4!}+\frac{D_{\Omega }}{2}\frac{%
c^{2}D_{B}t^{4}}{3!}\right. \notag \\
&&\left. +\frac{c^{2}D_{B}t^{3}}{2}+\frac{D_{\Omega }D_{B}^{2}t^{3}}{4}+%
\frac{D_{B}^{2}t^{2}}{2}\right],  \notag \\
&=&k^{4}f(t).   \label{pshort}
\end{eqnarray}
We now use Eq.~(\ref{forth}), that together with Eq.~(\ref{ms}), one finally gets for
short times that
\begin{equation}
\kappa \approx \frac{4^{3}f(t)}{\left( 4D_{B}t+2D_{B}D_{\Omega
}t^{2}+U_{0}^{2}t^{2}\right) ^{2}}=8,  \label{kure1}
\end{equation}
which means that our derived p.d.f. has a Gaussian behavior at short times.
This result has been validated numerically and illustrated in Fig.~\ref{Figkur}%
, where kurtosis is plotted versus time. It clearly can be seen that for
short times $\kappa $ has a value of $8$.

\subsection{Kurtosis at long times}

In ordet to find $\kappa $ for $t\rightarrow \infty ,$ we proceed as before, we expand 
$P\widetilde{P}_{0}(\boldsymbol{k},t)$ in power series of $k$, use that $\omega _{k}^{2}\equiv
U_{0}^{2}k^{2}/2-D_{\Omega }^{2}/4$, but now keep all the terms in the series expansion.
After applying Eq. (\ref{forth}) we get that in the long time limit 
\begin{eqnarray}
\langle \boldsymbol{x}^{4}(t)\rangle &\approx &\frac{2^{5}U_{0}^{4}}{%
D_{\Omega }^{4}}\left[ 1+\frac{4D_{B}D_{\Omega }}{U_{0}^{2}}\right.  \notag
\\
&&\left. +4\left( \frac{D_{B}D_{\Omega }}{U_{0}^{2}}\right) ^{2}\right]
\left( \frac{D_{\Omega }t}{2}\right) ^{2}.  \label{forthlim}
\end{eqnarray}
Finally we use Eq. (\ref{meanfree}) to get for long times that
\begin{equation}
\kappa =\frac{\langle \boldsymbol{x}^{4}(t)\rangle }{4\left( D_{B}+\frac{U_{0}^{2}}{%
2D_{\Omega }}\right) t}=8,  \label{kur2}
\end{equation}
which once again, Eq. (\ref{kur2}) shows that kurtosis has a Gaussian
behavior for long times. This result has also been validated using Brownian
dynamics and shown in Fig.~\ref{Figkur}, where for long values of the
dimensionless time, $\kappa$ tends again to a value of $8$.

\section{\label{sect_velfield}The active flux: the velocity field $\boldsymbol{V}(\boldsymbol{x},t)$ of active particles}
 
The Fourier expansion given by Eq.~\eqref{s1} allows us to obtain the hydrodynamic fields.
At the moment we have explicitly calculated $\widetilde{P}_{0}(\boldsymbol{k}
,t)$ that corresponds to the Fourier transform of the density field $\rho(
\boldsymbol{x},t)$. The Fourier transform of the components of the velocity 
field $\boldsymbol{V}(\boldsymbol{x},t)$ are related to $\widetilde{p}_{\pm 1}
(\boldsymbol{k},t)$ in the following way. By definition 
\begin{subequations}\label{Vfield}
\begin{align}
\widetilde{V}_{x}(\boldsymbol{k},t) &=U_{0}\int_{0}^{2\pi }d\varphi\, \cos \varphi 
\widetilde{P}(\boldsymbol{k},\varphi ,t) , \\
\widetilde{V}_{y}(\boldsymbol{k},t) &=U_{0}\int_{0}^{2\pi }d\varphi\, \sin \varphi 
\widetilde{P}(\boldsymbol{k},\varphi ,t),
\end{align}%
\end{subequations}
and after substitution of Eq.~\eqref{s1} in Eq.~\eqref{Vfield} we get
\begin{subequations}
\begin{align}
{\widetilde{V}_{x}}(\boldsymbol{k},t) =&\frac{U_{0}}{2}\, e^{-(D_{B}k^{2}+D_{\Omega})t}\, 
\left[\widetilde{p}_{1}(\boldsymbol{k},t)+\widetilde{p}_{1}^{*}(-\boldsymbol{k},t)\right] ,  \label{velx} \\
{\widetilde{V}_{y}}(\boldsymbol{k},t) =&\frac{U_{0}}{2}\, e^{-(D_{B}k^{2}+D_{\Omega})t}\, 
\left[\widetilde{p}_{1}^{*}(-\boldsymbol{k},t)-\widetilde{p}_{1}(\boldsymbol{k},t)\right],  \label{vely}
\end{align}
\end{subequations}
where we have used that $\widetilde{p}_{-n}(\boldsymbol{k},t)=\widetilde{p}_{n}^{*}(-\boldsymbol{k},t).$
With these expressions, it is straightforward to verify that Eq. \eqref{rec1} 
corresponds to the continuity equation $\partial_{t}\rho+\nabla\cdot\boldsymbol{V}(\boldsymbol{x},t)=0.$
At the long time regime, where only the first three Fourier modes are needed, 
we have that $\widetilde{p}_{1}(\boldsymbol{k},t)$ can be computed from Eq. (\ref{lead2}) if we
neglect the term that involves $p_{2},$ i.e., from
\begin{equation}\label{p1}
{\frac{d}{dt}\widetilde{p}_{1}}=-\frac{U_{0}}{2}ike^{-i\theta}\, e^{D_{\Omega
}t}\widetilde{p}_{0}.
\end{equation}
We now use the explicit expression for $\widetilde{p}_{0}$, that is Eq.~\eqref{lead2}, with
$\widetilde{p}_{0}(\boldsymbol{k},0)=1/(2\pi)$ to integrate 
Eq. \eqref{p1}. After some steps we find that
\begin{equation}
\widetilde{p}_{1}(\boldsymbol{k},t)=-\frac{U_{0}}{4\pi}\left(
ik_{x}+k_{y}\right)e^{
D_{\Omega}t/2}\frac{\sin \omega _{k}t}{\omega _{k}}.  \label{proba1}
\end{equation}%
If we insert the latter into Eqs.~(\ref{velx})-(\ref{vely}) we obtain 
\begin{eqnarray*}
\widetilde{V}_{x}(\boldsymbol{k},t) &=&-\frac{U_{0}^{2}}{4\pi}ik_{x}
e^{-\left( D_{B}k^{2}+D_{\Omega }/2\right)t}\frac{\sin
\omega _{k}t}{\omega _{k}}, \\
\widetilde{V}_{y}(\boldsymbol{k},t) &=&-\frac{U_{0}^{2}}{4\pi}ik_{y}
e^{-\left( D_{B}k^{2}+D_{\Omega }/2\right)
t}\frac{\sin
\omega _{k}t}{\omega _{k}}.
\end{eqnarray*}
After inverting the Fourier transform, this pair of expressions can be brought 
into the form
\begin{equation}
 \boldsymbol{V}(\boldsymbol{x},t)=-\frac{U_{0}^{2}}{2}\nabla F(\boldsymbol{x},t),
\end{equation}
where the function $F(\boldsymbol{x},t)$ depends on $P_{0}$, explicitly
\begin{equation}
 F(\boldsymbol{x},t)=\int_{0}^{t}ds\, e^{D_{\Omega}(t-s)}\int d^{2}\boldsymbol{x}^{\prime}
 G(\boldsymbol{x}-\boldsymbol{x}^{\prime},t-s)P_{0}(\boldsymbol{x}^{\prime},s).
\end{equation}
Last expression corresponds to a non-Fickian constitutive relation that considers
nonlocal effect in space and time. This relation, together with the continuity equation,
gives rise to a non-local (in both, space and time) diffusion-like equation.

In a similar way, one can systematically find higher Fourier modes for the probability
density function, $\widetilde{P}({\boldsymbol{k}},\varphi ,t)$, in the
wave vector space.

\section{Final Comments and Conclusions}

Though Eq.~\eqref{MSD_analytical} has been derived from the long time
approximation, it does give the correct time dependence at all times as it is
shown in Fig. \ref{Fig_msd}, where our analytical formula (solid lines) is
compared against numerical simulations (dashed lines). This fact can be
understood directly from Eq.~\eqref{lead0}, since it can be shown that
the inhomogeneous term does not contribute to the mean squared displacement
(see appendix \ref{appendixA}), therefore agreeing with the expression obtained from the
telegrapher's Eq.~\eqref{lead}.

\begin{figure}
 \includegraphics[width=\columnwidth]{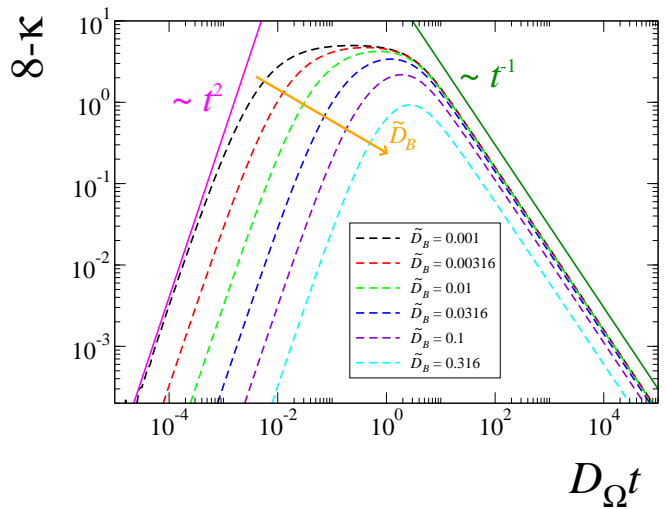}
 \caption{(Color online) The departure from Gaussian behavior of $P_{0}(\boldsymbol{x},t),$
 measured by $8-\kappa$, differs in the short-time regime from the long-time regime. In
 the first case the departure captured by the power law $(D_{\Omega}t)^{-2}$ while in the 
 long-time regime such departure is well described simply by $(D_{\Omega}t)^{-1}$. }
 \label{Fig5}
\end{figure}

We want to point out that though the linear time dependence of the msd,
characteristic of normal diffusion, is reached at times $D_{\Omega}t \sim1$
(see Fig. \ref{Fig_msd}), the Gaussian behavior of the distribution is
reached at much later times (see Fig. \ref{Figkur}), suggesting that at
finite times, $D_{\Omega}t>1$, normal diffusion does strictly corresponds
to Gaussian distributions.
The kurtosis computed from our approximation
shows that it tends asymptotically to Gaussian, $\kappa=8$, as $%
(8-D_{\Omega}t)^{-1}$. In contrast, in the short time regime, the
distribution departs from a Gaussian behavior with time as $(8-D_{\Omega}t)^{-2}$ (see Fig. \ref{Fig5}).

In summary, we have found a method and obtained with it, an analytical solution to the
Smoluchowski equation of an active Brownian particle self-propelling with a constant velocity.
We used a Fourier approach and exploited the circular symmetry of the distribution function,
to obtain an infinite
system of coupled ordinary differential equations for the Fourier modes of the probability
density. Our formalism showed that in order to have a whole description for the particles 
diffusion, we only need the first Fourier coefficient, since it was found that higher Fourier
modes do not contribute to the mean-square displacement of the particle.  With the explicit
p.d.f. in hand, we calculated the particle mean-square displacement for both long and short
times, recovering classical results of enhanced diffusion due to activity. We also validated
these findings by performing Brownian dynamics simulations that showed and excellent agreement
among theory and simulations. In addition, we also performed an asymptotic analysis for short
and long times, to the forth moment of our p.d.f. This calculation enabled us to have analytical
results for the kurtosis. The asymptotic results for this measure were validated with Brownian
dynamics simulations that showed that the p.d.f. of an active particle always starts with a
Gaussian behavior, and depending on the vale of the ratio $\widetilde{D}_{B}$, it may
evolve to a ring-like distribution, but it always returns to a Gaussian distribution
for long times.  Finally this work  shed light on mathematical procedures to solve a
Smoluchowski equation for an active Brownian particle.

Future venues for this research, would be to solve analytically a Smoluchowski equation
for confined active particles  and/or particles interacting among them.

\begin{acknowledgments}
 F.J.S. acknowledges support from DGAPA-UNAM through
Grant No. PAPIIT-IN113114. M. S. thanks CONACyT
 and Programa de Mejoramiento de Profesorado (PROMEP) for partially funding this work.
\end{acknowledgments}

\appendix
\section{\label{appendixA} Mean Square Displacement}

If we apply the operator $-\nabla^{2}_{\boldsymbol{k}}$ 
to the relation $\widetilde{P}_{0}(\boldsymbol{k},t)=e^{-D_{B}{k}^{2}t}\widetilde{p}%
_{0}(\boldsymbol{k},t)$ and evaluate at $\boldsymbol{k}=0$, one gets that 
$\langle\boldsymbol{x}^{2}(t)\rangle=4D_{B}t+
\langle\boldsymbol{x}^{2}(t)
\rangle_{0}$, which leads straightforwardly to the pair of relations
\begin{subequations}\label{rels}
 \begin{align}
 \frac{d^{2}}{dt^{2}}\langle\boldsymbol{x}^{2}(t)\rangle&=
\frac{d^{2}}{dt^{2}}\langle\boldsymbol{x}^{2}(t)
\rangle_{0} \\
\frac{d}{dt}\langle\boldsymbol{x}^{2}(t)\rangle&=
4D_{B}+\frac{d}{dt}\langle\boldsymbol{x}^{2}(t)
\rangle_{0},
 \end{align}
\end{subequations}
where $\langle\cdot\rangle_{0}$ denotes the average of $(\cdot)$ with $p_{0}
(\boldsymbol{x},t)$ as the probability measure. The quantity $\frac{d^{2}}{dt^{2}}
\langle\boldsymbol{x}^{2}(t) \rangle_{0}$ can be computed directly by applying $-\nabla^{2}_{\boldsymbol{k}}$
to Eq.~\eqref{lead0} and evaluating at $\boldsymbol{k}=0$, which results in
\begin{multline}
\frac{d^{2}}{dt^{2}}\langle\boldsymbol{x}^{2}(t)\rangle_{0}+D_{\Omega}\frac{d}{dt}\langle\boldsymbol{x}^{2}(t)\rangle_{0}=\frac{U_{0}^{2}}{2}\left[\nabla^{2}_{\boldsymbol{k}}(k_{x}^{2}+k_{y}^{2})\widetilde{p}_{0}\right]_{\boldsymbol{k}=0}\\
+ \frac{U_{0}^{2}}{4}e^{-4D_{\Omega}t}\left\{\nabla^{2}_{\boldsymbol{k}}\left[(k_{x}-ik_{y})^{2}\widetilde{p}_{-2}+(k_{x}+
 ik_{y})^{2}\widetilde{p}_{2}\right]\right\}_{\boldsymbol{k}=0}
\end{multline}
It is easy to show that the first term of rhs of the last equation 
is $2U_{0}^{2}$ while the next term, that involves the modes $\widetilde{p}_{\pm2}$ 
vanishes identically,
thus we get 
\begin{equation}
\frac{d^{2}}{dt^{2}}\langle\boldsymbol{x}^{2}(t)\rangle_{0}+D_{\Omega}\frac{d}{dt}
\langle\boldsymbol{x}^{2}(t)\rangle_{0}=2U_{0}^{2},
\end{equation}
and therefore, using Eqs. \eqref{rels} we finally have that
\begin{equation}
\frac{d^{2}}{dt^{2}}\langle\boldsymbol{x}^{2}(t)\rangle+D_{\Omega}\frac{d}{dt}
\langle\boldsymbol{x}^{2}(t)\rangle_{0}=2U_{0}^{2}+4D_{B}D_{\Omega},
\end{equation}
which coincides with Eq. \eqref{ode}, deduced from the telegrapher Eq.~\eqref{lead}
when considering only the first three Fourier modes $\widetilde{p}_{0}\, ,\widetilde{p}_{\pm1}$.

\section{\label{appendixB} Calculation of the Fourth moment of $P_{0}(\boldsymbol{x},t)$}

The fourth moment $\langle\boldsymbol{x}^{4}(t)\rangle_{r}$ can be obtained through 
the prescription 
\begin{equation}
\langle \boldsymbol{x}^{4}(t)\rangle =\left( \frac{1}{k}\frac{\partial }{%
\partial k}k\frac{\partial }{\partial k}\right)^{2} \left. \widetilde{%
P}_{0}(\boldsymbol{k},t)\right\vert _{k=0}.  \label{forth}
\end{equation}
Though it is possible to use this prescription directly on $\widetilde{P}_{0}(\boldsymbol{k},t)$,
we prefer to apply it to the series expansion in powers of $k$ of $\widetilde{P}_{0}(\boldsymbol{k},t)$,
this is due to the complicated dependence of $\widetilde{P}_{0}$ on $k$. Such expansion is given by
\begin{multline}
\widetilde{P}_{0}(\boldsymbol{k},t)=e^{-D_{\Omega}t/2}\sum_{n,m=0}^{\infty}\sum_{j=0}^{m}\binom{m}{j}\frac{(-D_{B}t)^{n}}{n!}
\frac{(it)^{2m}}{(2m)!}\\
\times \left(1+\frac{D_{\Omega}t}{2(2m+1)}\right)\left(-\frac{D_{\Omega}^{2}}{4}\right)^{m-j}\left(\frac{U_{0}^{2}}{2}\right)^{j}\, k^{2(n+j)}.
\end{multline}
where we have used the explicit dependence on $k$, of $\omega_{k},$ given in section \ref{SectModel}.

Since 
\begin{multline}
 \left( \frac{1}{k}\frac{\partial }{%
\partial k}k\frac{\partial }{\partial k}\right)^{2}k^{2(n+j)}=\left[2(n+j)\right]^{2}\times\\
\left[2(n+j-1)\right]^{2}k^{2(n+j-2)},
\end{multline}
the only terms that survives when evaluating at $k=0$ are those for which $n+j=2$ with the condition 
that $0\le j\le m$. Thus we can write
\begin{widetext}
\begin{multline}
 \langle \boldsymbol{x}^{4}(t)\rangle =4^{4}\frac{U_{0}^{4}}{D_{\Omega}^{4}}e^{-D_{\Omega}t/2}
 \sum_{m=0}^{\infty}\left(\frac{D_{\Omega}t}{2}\right)^{2m+2}\left[
 \left(\frac{D_{\Omega}t}{2}\right)^{2} \left(1+\frac{D_{\Omega}t}{2}\frac{1}{2m+5}\right)\frac{(m+2)(m+1)}{2(2m+4)!}+\right.\\
 \widetilde{D}_{B}\left(\frac{D_{\Omega}t}{2}\right) \left(1+\frac{D_{\Omega}t}{2}\frac{1}{2m+3}\right)\frac{m+1}{(2m+2)!}+
 \widetilde{D}_{B}^{2} \left.\left(1+\frac{D_{\Omega}t}{2}\frac{1}{2m+1}\right)\frac{1}{2(2m)!}\right]
\end{multline}
\end{widetext}
The sums can be written in terms of hyperbolic function, and after some algebraic steps Eq.~\eqref{4thMoment} is recovered.


\providecommand{\noopsort}[1]{}\providecommand{\singleletter}[1]{#1}%

\end{document}